# Structural Design and Performance Analysis of Laser Transmitting Telescope for Space Gravitational Wave Detection

Long Yongtao[1], Mo Yan[2], Cao Shengyi[1], Cao Jiamin[2], Zhao Lujia[2], Wang Haibo[2], Wang Shuangbao[1], Tan Hao[2], Liu Xiaohong[3], Wang Dawei[3], Ma Donglin[1,2,*]

1 School of Optical and Electronic Information, Huazhong University of Science and Technology, Wuhan 430074, China;

2 School of Physics, Huazhong University of Science and Technology, Wuhan 430074, China;

3 Optics Valley Laboratory, Wuhan 430074, China

**Abstract**: The spaceborne laser emission telescope is a core and critical component of the space gravitational wave detection system. Compared with ground-based telescopes, the on-orbit space environment is more complex and harsh, presenting higher technical challenges for the design of the optical system and structure — both optical design and structural design face considerable difficulties. To meet the requirements of space gravitational wave detection, this paper designs a laser emission telescope based on an off-axis four-mirror configuration, with a capture field of view of ±300μrad, an optical transmission efficiency of 86.3%, and an optical path stability index of TTL≤0.025 nm/μrad. During the design process, based on existing theories and engineering experience, the primary mirror thickness optimization and lightweight structural design were completed, and a flexible support scheme was adopted to achieve a primary mirror surface figure accuracy of 9.42 nm; the total mass of the entire telescope (excluding mirrors) is only 3.845 kg. Multi-dimensional finite element analysis was conducted on the telescope under actual working conditions: the strength of the telescope's support materials was verified under self-weight and 10G gravity loads; after removing the rigid body displacement of the mirrors using Zernike polynomials, the surface deformation of the primary mirror was controlled within 1/30 wavelength. In the thermal stability analysis, the structural deformation of the telescope under a temperature change of 100°C was simulated, and key indicators such as eccentricity and tilt between the mirrors all meet the optical design requirements. In the modal analysis, the first-order natural frequency of the telescope reaches 200 Hz under both self-weight and weightless conditions, demonstrating excellent dynamic stability. The research results indicate that both the optical performance and structural reliability of the telescope meet the operational requirements for space gravitational wave detection.

## 1  Introduction

Gravitational waves, as the propagation form of spacetime perturbations in the universe, are a key prediction of general relativity. Their detection provides a new perspective for revealing scientific issues such as cosmic evolution and the interaction of compact objects (black holes, neutron stars). Since the first direct detection of the gravitational wave signal from the merger of binary black holes (GW150914) by the US Laser Interferometer Gravitational-Wave Observatory (LIGO) in 2015 [1], more than 90 gravitational wave events have been discovered globally through ground-based interferometers (LIGO, Virgo, KAGRA) [2]. However, limited by noises such as ground vibration and gravity gradient, ground-based facilities cannot cover the low-frequency gravitational wave band below 1 Hz. This band corresponds to important wave sources including white dwarf binaries and massive black hole mergers, which must be detected by spaceborne interferometers [3]. As a core component of spaceborne interferometers, the surface figure accuracy of the reflective mirrors and the stability of the support structure of spaceborne laser transmitting telescopes directly determine the laser ranging accuracy, thereby affecting the sensitivity of gravitational wave detection. Therefore, the high-precision structural design of spaceborne telescopes has become a key technical bottleneck for space gravitational wave detection missions.

Internationally, the Laser Interferometer Space Antenna (LISA) project jointly promoted by the European Space Agency (ESA) and the National Aeronautics and Space Administration (NASA) is a representative program for low-frequency gravitational wave detection. It adopts a three-satellite heliocentric orbital formation with an arm length of 5 million km. The LISA Pathfinder satellite launched in 2015 verified key technologies such as drag-free control and ultra-stable optical platforms, but was not equipped with a spaceborne laser transmitting telescope [4]. The off-axis telescope prototype developed by the Netherlands Organisation for Applied Scientific Research (TNO) in 2012 met the natural frequency index but failed to satisfy the structural stability requirements [5]. NASA developed a coaxial all-silicon carbide telescope prototype and an off-axis four-mirror prototype in 2012 and 2016, respectively. The former failed to meet performance standards at -60 ℃ due to temperature fluctuations [6], while the latter achieved a wavefront stability of 24 nm at room temperature but without extreme environment verification [7]. In China, two major space gravitational wave detection schemes have been formed: the Taiji Program and the Tianqin Program. The Taiji Program, led by the Chinese Academy of Sciences, adopts a three-satellite heliocentric formation with an arm length of 3 million km. Li Yupeng, Wang Chenzhong, et al. designed 216 mm and 210 mm aperture primary mirror support structures respectively, with an on-orbit surface figure accuracy of 8.83–8.9 nm [8–9]. The Tianqin Program, proposed by Academician Luo Jun in 2014 and led by Sun Yat-sen University, adopts a three-satellite geocentric orbital formation with an arm length of 170,000 km. The Tianqin-1 technology satellite was successfully launched in 2019 and completed global gravity field measurement [10–11]. However, current research on spaceborne telescopes for the Tianqin Program still suffers from problems such as unclear structural design indicators, lack of primary mirror lightweight parameters, and absence of an integrated four-mirror support scheme. A complete structural design and performance verification scheme is urgently needed.

This paper mainly conducts a collaborative design study on the optical system and structural system of space gravitational wave detection telescopes. Firstly, the optical system design of the telescope is completed, which serves as the core foundation for the subsequent structural design and defines the tolerance control standards for structural design. On this basis, the support structures of each mirror are designed. For large-aperture primary mirrors, a lightweight design scheme is adopted to balance structural strength and mass requirements. For other mirrors, high-precision fine-tuning mechanisms are designed to ensure mirror position accuracy. Finally, the performance of the whole telescope is verified through finite element analysis, with emphasis on evaluating the structural stability under the Earth's gravity environment and 10-gravity load during transportation. The influence of temperature variation on mirror spacing is analyzed using Ansys software, and the modal characteristics of the whole machine are studied. The results show that the designed structural system can fully meet the design indicators of the optical system and can provide a valuable reference for the design of similar space gravitational wave detection telescopes

## 2 Optical Design of Laser Transmitting Telescope for Space Gravitational Wave Detection

Compared with traditional telescope systems, a significant difference of the laser transmitting telescope for space gravitational wave detection is that it operates in a bidirectional manner. While performing laser transmission, it also needs to receive signal light emitted by distant gravitational wave detection telescopes. Therefore, to ensure the optical path stability of the telescope system, extremely stringent requirements are imposed on wavefront aberrations to avoid introducing unnecessary additional jitter noise during subsequent interferometric measurements. During the optical design of the telescope, system optimization is required to achieve small residual aberrations within the field of view [12–13].

The laser transmitting telescope for space gravitational wave detection is planned to operate in Earth orbit, with three independent spacecraft forming an equilateral triangular formation. Consequently, the telescope imposes high requirements on both imaging quality and structural design. In this paper, an off-axis four-mirror configuration is adopted for the optical design of the telescope. Compared with conventional coaxial systems, the off-axis system can further reduce the influence of stray light while ensuring favorable imaging quality [14]. In the optimization process, to further improve the aberration tolerance of the system, even-order aspheric surfaces are introduced into the system in addition to off-axis mirror configuration, which enhances the aberration correction capability of the secondary mirror, tertiary mirror and quaternary mirror. The detailed optical layout of the laser transmitting telescope for space gravitational wave detection is shown in Fig. 1.

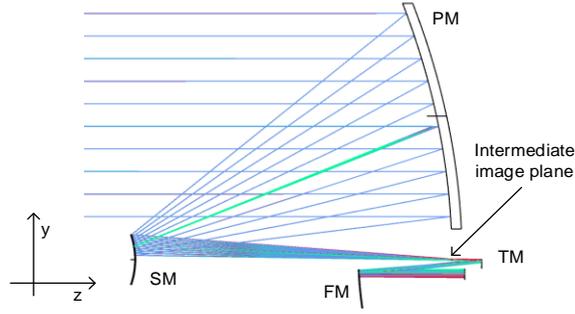

Fig. 1 Optical Layout Diagram of the Laser Emitting Telescope for Space Gravitational Wave Detection

The telescope system operates at a wavelength of 1064 nm with a capture field of view of ±300 μrad. At this wavelength, the single-surface reflectivity of the telescope is approximately 99%, resulting in an overall system transmission efficiency of about 96% [15]. During the design process, optimization was focused on the RMS wavefront error within the capture field of view. The optimized RMS wavefront error curve of the system is shown in Fig. 2. The results indicate that the system exhibits a relatively uniform wavefront distribution across the capture field of view, and the RMS wavefront error in all fields of view is less than λ/300, providing an optical design model and sufficient design margin for the subsequent structural design of the system.

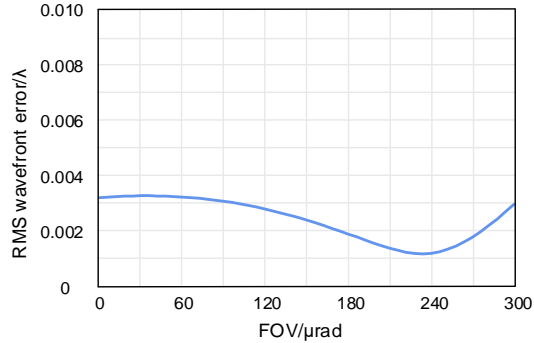

Fig. 2 RMS Wavefront Error Distribution Diagram of the Laser Emitting Telescope for Space Gravitational Wave Detection

Furthermore, in the space environment, the system will inevitably be affected by field of view jitter. The angular misalignment introduced by such jitter will lead to errors in the measurement phase, which is generally referred to as jitter-induced optical path coupling noise. Therefore, after completing the optical design of the telescope system, the TTL coupling noise is simulated and analyzed with reference to the simulation models in existing studies [12–14], and the simulation results are shown in Fig. 3. The results show that within the angular misalignment range of ±300 μrad, the maximum TTL noise of the system is approximately 0.015 nm/μrad, which meets the design requirements of optical path stability.

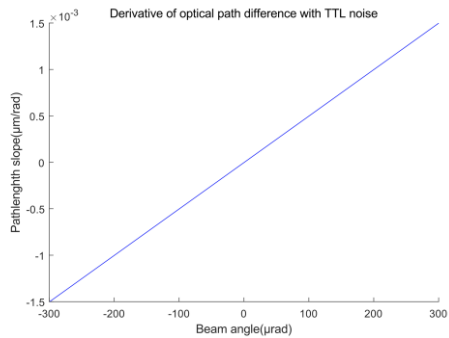

Fig. 3 Curve of slope of pathlength signal with tilt angle

According to the telescope design results, the detailed optical design specifications of the system are shown in Table 1.

Table 1 Optical Specifications of the Laser Emitting Telescope for Space Gravitational Wave Detection

| Parameters | Design Requirements |
|---|---|
| Wavelength | 1064nm |
| Designed Participating Aberration | ≤λ/300 RMS@1064 nm |
| Capture Field of View | ±300μrad |
| Science Field of View | ±7μrad |
| Entrance Pupil Diameter | 220nm |
| Magnification | 40× |
| Optical Transmission Efficiency | 86.3% |
| Optical Path Stability | TTL≤0.025 nm/μrad |

# 3  Structural Design of Laser Transmitting Telescope for Space Gravitational Wave Detection

The support structure of the off-axis four-mirror spaceborne laser transmitting telescope designed in this paper is mainly composed of a primary mirror support module, a secondary mirror support module, a tertiary mirror support module, a quaternary mirror support module, and inter-module connecting structures. The primary mirror has an aperture of 220 mm, while the apertures of the other mirrors are all no more than 50 mm. Lightweight design is adopted for the primary mirror to reduce its self-weight, and corresponding fine-adjustment mechanisms are designed for the other mirrors to ensure high-precision spatial positioning.

## 3.1  Lightweight and Support Structure Design of the Primary Mirror

During the operation of the telescope, the ambient temperature varies considerably. Meanwhile, to mitigate the influence of gravity load on the surface figure accuracy of the primary mirror, lightweight design and flexible support design are required for the primary mirror. Firstly, the structural parameters of the primary mirror are estimated based on empirical formulas, and then a three-dimensional model is constructed according to these parameters. The surface deformation of the mirror is solved by means of finite element analysis. Finally, combined with the deformation analysis results, multi-dimensional dimensional optimization is carried out for the inner diameter, rib thickness, and minimum thickness of the mirror.

The structural parameters mainly determine the thickness $T$ of the mirror, the mirror surface thickness $t_f$, and the radius $s$ of the lightweight region, as shown in Fig. 4. According to the empirical formula for mirror thickness proposed by Robert et al[16].

$$\delta = \frac{3\rho g a^4}{16 E T^2} \quad (1)$$

where $\delta$ represents the maximum deformation of the mirror, $a$ is the mechanical half-diameter of the mirror, $\rho$ is the density of the mirror material, $g$ denotes the gravitational acceleration, $E$ is the Young's modulus of the mirror material, and t stands for the thickness of the mirror. The mirror designed in this paper uses Zerodur glass-ceramic as the material, whose material parameters are listed in the table. The design requirement is that the RMS surface error of the mirror under 1G gravity is less than 10 nm, and the mechanical half-diameter of the mirror is 110 mm. Substituting the parameters into the formula, the minimum thickness of the mirror is calculated to be 27.2 mm.

For the convenience of subsequent calculation and manufacturing, the thickness T is initially set to 27.5 mm.

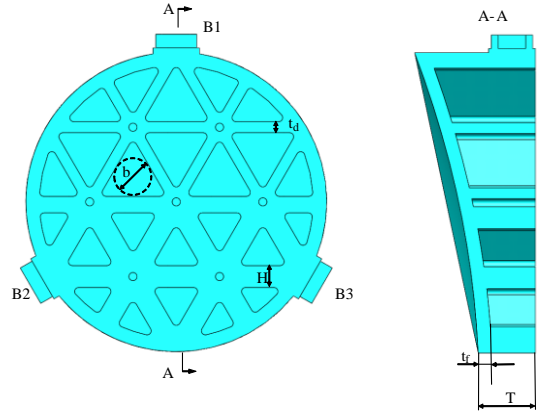

Fig. 4 Schematic Diagram of Primary Mirror Dimensions

The maximum deformation of a single honeycomb aperture on the mirror surface can be expressed according to the deduction by Barnes [17] as

$$\delta = \psi\left(\frac{P}{D}\right)b^4 = \psi\left[\frac{Et_f^3}{12(1-v^2)}\right]^{-1} Pb^4 \qquad (2)$$

where $t_f$ represents the thickness of the mirror surface, $P$ is the uniform positive pressure applied on the honeycomb cells during mirror processing, b is the inscribed circle diameter of the triangular lightweight honeycomb aperture, and $\Psi$ is the shape factor of the triangle. In this paper, $P$ is taken as 67 kPa, b is taken as 30 mm, and the value of $\Psi$ is 0.00151. It can be seen that, under a given deformation, the cube of the back rib thickness of the mirror is proportional to the fourth power of the inscribed circle diameter b of the lightweight aperture. According to the design requirement that the deformation is less than 10 nm, the primary mirror surface thickness is calculated to be 10.024 mm, and $t_f$=10 mm is adopted.

The primary mirror support structure adopts a BIPOD flexure hinge design. When subjected to external stress, the flexure hinge can produce elastic deformation, thereby reducing the influence of structural deformation on the primary mirror surface. The dimensional parameters of the flexure hinge are shown in Fig. 5. The primary mirror adopts a side-support configuration, and counterweight compensation is applied via other structures to mitigate the deformation caused by the mirror's self-weight, as illustrated in Fig. 6.

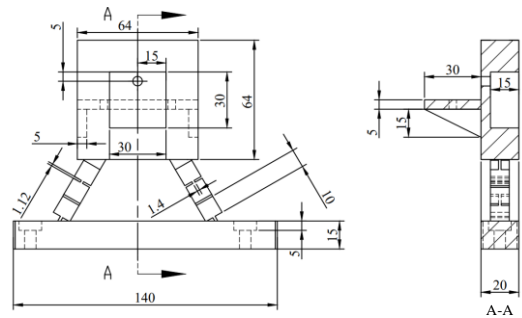

Fig. 5 Schematic Diagram of Flexure Hinge Dimensions

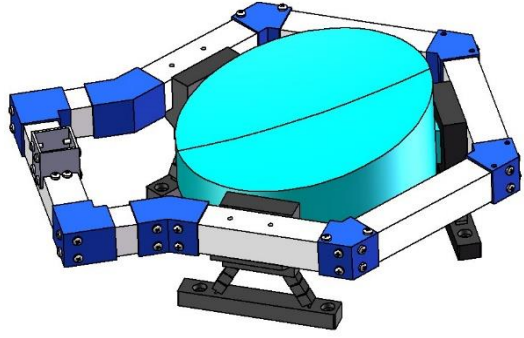

Fig. 6 Primary Mirror Support Scheme

After determining the dimensional parameters and support scheme of the mirror, multi-dimensional optimization can be carried out according to the finite element analysis results of the mirror surface. Optimizations are mainly performed on the inner diameter $s$, rib thickness $t_d$, and minimum thickness $T$ of the mirror as shown in Fig. 3. The parameter constraints of the mirror are as follows

$$\begin{cases} s_0 = 103mm \\ t_0 = 6mm \\ T_0 = 27.5mm \\ 100 \leq s \leq 106 \\ 4 \leq t_d \leq 10 \\ 26.5 \leq T \leq 55 \end{cases} \quad (3)$$

where $s_0$, $t_0$ and $T_0$ are the initial values of the three parameters. Since the computation of RMS is more intensive than that of PV, the PV value is set as the optimization objective. The preliminary optimization goal is to minimize the mass of the mirror while satisfying PV < 40 nm (four times the 10 nm RMS design requirement). The step size of $t_d$ and $T$ is set to 1 mm, and the step size of $s$ is set to 0.5 mm. A total of 27 optimization iterations are carried out. The gravity simulation results of the optimized primary mirror are shown in Fig. 6. When gravitational acceleration is applied in different directions, the RMS values of the optimized primary mirror surface are 7.56 nm, 3.88 nm and 9.42 nm, respectively. The surface figure in all directions meets the optical design requirements, and the dimensional parameters are listed in Table 2.

Table 2 Parameters and Dimensions of the Primary Mirror

| Parameters | Value/mm |
|---|---|
| $T$ | 41.5 |
| $t_f$ | 10 |
| $L$ | 3 |
| $H$ | 16 |
| $h$ | 8 |
| $b$ | 25.83 |
| $s$ | 101.5 |

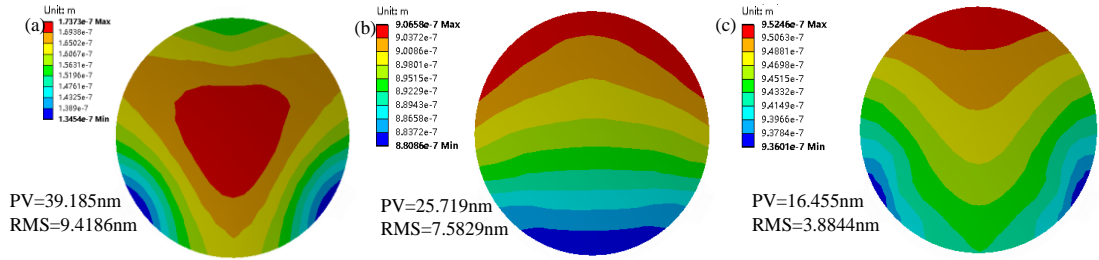

<div align="center">
PV=39.185nm  PV=25.719nm  PV=16.455nm
RMS=9.4186nm  RMS=7.5829nm  RMS=3.8844nm
</div>

Fig. 7 Gravity Simulation Results of the Primary Mirror in All Directions After Optimization: (a) Gravitational Deformation of the Primary Mirror in the Z-axis Direction; (b) Gravitational Deformation of the Primary Mirror in the X-axis Direction; (c) Gravitational Deformation of the Primary Mirror in the Y-axis Direction

**3.2 Design of Support Structures for the Other Three Mirrors**

In the design of support structures for the secondary, tertiary, and quaternary mirrors, emphasis is placed on the influence of external temperature fluctuations and assembly stress on surface figure accuracy. A flexible structure design is adopted to achieve stress release, and a fine-adjustment mechanism is designed to facilitate subsequent alignment and ensure that the spatial positioning accuracy of each mirror meets the design requirements.

The secondary mirror uses Zerodur glass-ceramic as the substrate material, with an aperture of 48.8 mm and a thickness of 1.5 mm. Lightweighting is unnecessary due to its small mass. The support structure of the secondary mirror adopts a flexible design. The back of the mirror is bonded to a precision-ground pad, and a flexible gap is arranged between the pad and the fixed outer ring, which can effectively isolate the deformation of the support structure caused by temperature variation from being transmitted to the mirror and significantly improve the temperature adaptability of the structure. Temperature deformation analysis was carried out for the conventional rigid support structure and the flexible support structure. The results are shown in Fig. 8. The PV value of the mirror is reduced from 314 nm to 155 nm, with an improvement of 50.67%, which verifies the effectiveness of the flexible structure in improving the temperature adaptability of the mirror. Aiming at problems such as relative distance offset, lateral displacement, and tilt error between mirrors that may occur during the optical system alignment stage, a five-dimensional precision adjustment mechanism is configured for the secondary mirror: the mirror tilt is fine-tuned via tilt adjustment screws, and precise spatial positioning is realized by XY adjustment screws and Z-axis adjustment nuts. The overall support structure is shown in Fig. 9.

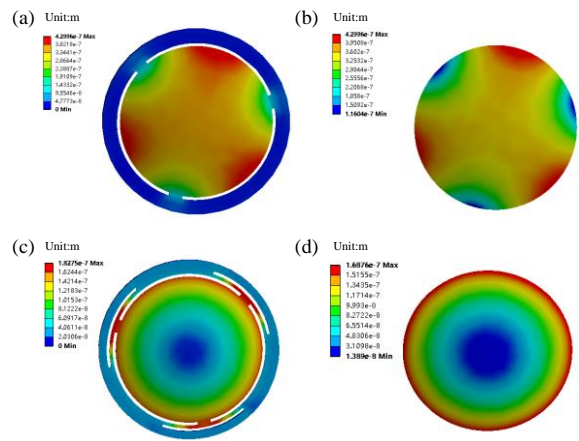

Fig. 8 Heat Deformation Results Under Different Supports: (a) Overall Deformation Under Ordinary Supports; (b) Mirror Surface Deformation Under Ordinary Supports; (c) Overall Deformation Under Flexible Supports; (d) Mirror Surface Deformation Under Flexible Supports

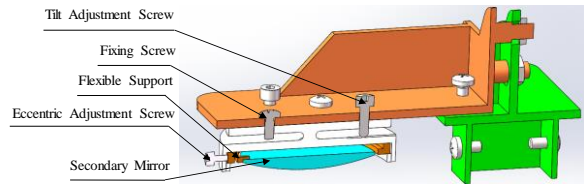

Fig. 9 Secondary Mirror Support Structure

The tertiary mirror has a clear aperture of 11 mm and a thickness of 2 mm. Given its small size, a flexible support structure is not required, and only a precision fine-adjustment mechanism is equipped to meet subsequent optical alignment requirements. Different from the secondary mirror support structure, the tertiary mirror support is firmly connected to the lower mirror mounting plate via fixing screws to form an integrated assembly module. This module can achieve precise fine adjustment of the mirror height through the Z-axis adjustment screws on the support base. After position calibration, Z-axis locking screws are used to fix the assembly and ensure the stability of its spatial position.

The quaternary mirror has an aperture of 37.8 mm and a thickness of 5 mm. A flexible structure is also designed for the quaternary mirror support to counteract stress induced by temperature variations, and a five-dimensional adjustment mount is adopted to facilitate alignment. The detailed structures of the tertiary mirror and quaternary mirror are shown in Fig. 10 and Fig. 11, respectively.

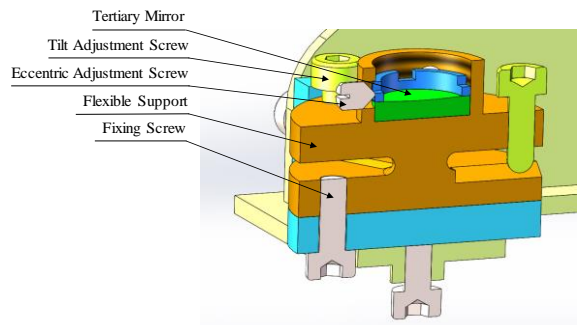

Fig. 10 Tertiary Mirror Support Structure

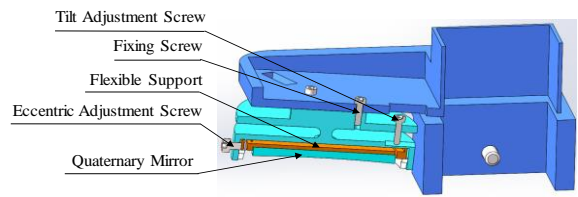

Fig. 11 Quaternary Mirror Support Structure

### 3.3 Design of Inter-Mirror Support Structure

The design of the inter-mirror connection structure needs to focus on two core indicators: structural stiffness and thermal expansion stability. The overall structure adopts a composite configuration of "hexagonal main body + rectangular outer frame". The primary mirror is mounted at the center of the regular hexagonal area, and the specific structure is shown in Fig. 12. The connection structure is manufactured with carbon fiber reinforced polymer (CFRP). The total mass of the support structure excluding all mirrors is 3.845 kg, which meets the preset mass requirement.

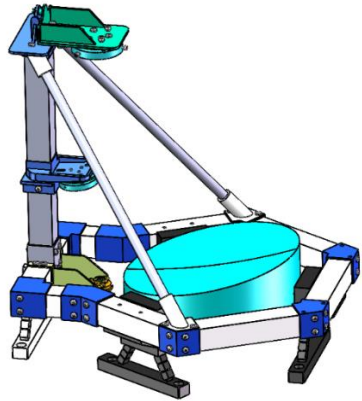

Fig. 11 Overall Structural Design

# 4 Finite Element Analysis of the Overall System

After completing the overall structural design, Ansys finite element analysis software is used to perform multi-dimensional simulation to systematically verify its mechanical reliability, thermal environment adaptability and dynamic characteristics. During ground assembly and testing, the telescope is mainly affected by gravity deformation, while during launch and transportation to orbit, it is subjected to 7–8 times the Earth's gravity. In addition, temperature variations in space can reach up to 100 °C, which will cause considerable changes in mirror spacing. Therefore, various environmental simulations are required to ensure that the structural strength meets the design thresholds. Finally, modal analysis is carried out on the whole telescope to obtain the natural frequencies and mode shapes, so as to avoid resonance within the environmental frequency range. Meanwhile, the Zernike polynomial is used to quantitatively decompose the surface error during simulation and evaluate the optical performance. By correlating the structural mechanical response with the optical imaging quality, the comprehensive reliability of the designed structure is fully verified, providing strong theoretical support for subsequent engineering applications.

**4.1 Static Analysis**

The assembly and alignment of the spaceborne laser transmitting telescope must be completed on the ground before being launched into space orbit by a carrier rocket. During ground assembly, the telescope bears the load of Earth's gravitational acceleration, while it operates in a weightless environment on orbit. Such a significant difference in gravity environment may lead to the spatial position deviation and surface figure error of the mirrors. Therefore, it is necessary to conduct static analysis related to gravity load on the telescope to guarantee the stability of its on-orbit optical performance.

The deformation characteristics of the overall structure under self-weight are analyzed by finite element simulation, and the results are shown in Fig. 13. It is found that the surface figure accuracy of the primary mirror is the worst when the gravity direction is along the X-axis, mainly due to the asymmetric weight distribution on both sides caused by the non-rotationally symmetric structure of the mirror. Accordingly, during ground assembly, the gravity direction should be aligned with the symmetry plane of the primary mirror, and it is recommended to preferentially apply the gravity load along the Z-axis or Y-axis.

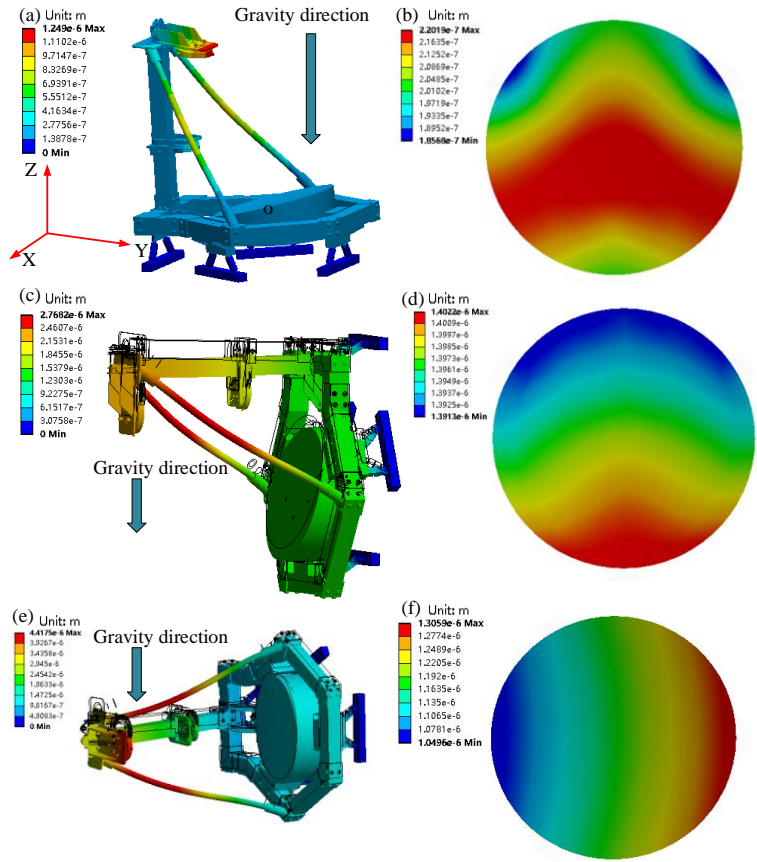

Fig. 13 Finite Element Analysis of Self-weight Deformation: (a) Gravitational Deformation of the Overall Structure in the Z-axis Direction; (b) Gravitational Deformation of the Primary Mirror in the Z-axis Direction; (c) Gravitational Deformation of the Overall Structure in the Y-axis Direction; (d) Gravitational Deformation of the Primary Mirror in the Y-axis Direction; (e) Gravitational Deformation of the Overall Structure in the X-axis Direction; (f) Gravitational Deformation of the Primary Mirror in the X-axis Direction

To verify the surface figure accuracy under gravity, the surface deformation of the primary mirror along the Z-axis direction is fitted using Zernike polynomials. Considering that the secondary, tertiary, and quaternary mirrors are equipped with precision adjustment mechanisms, the tilt deformation of the primary mirror can be compensated during ground alignment. Thus, the first three tilt-related terms of the Zernike polynomials are removed. The calculated RMS value of the primary mirror surface is 6.18 nm, as shown in Fig. 14, which meets the optical design requirement of 10 nm RMS.

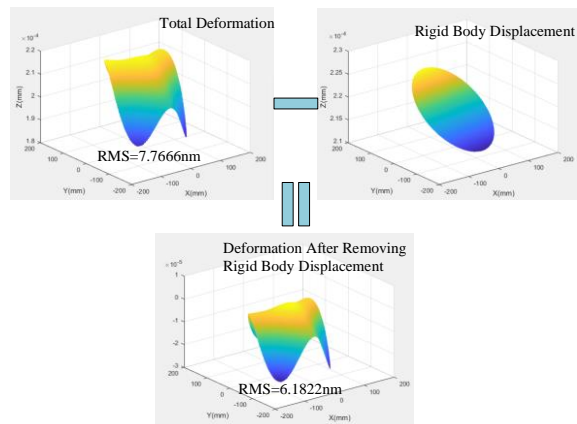

Fig. 14 Primary Mirror Surface After Removing Rigid Body Displacement

During the transportation of the telescope from the ground to orbit, it is subjected to a load of approximately 7–8 times the gravitational acceleration. To ensure that the structure does not fracture or suffer damage during this phase and that the deformation is fully recoverable, stress analysis of the overall structure under extreme acceleration is required. A simulated acceleration of 10G is applied to the telescope along the Z-axis and Y-axis directions respectively, and the structural stress nephograms are shown in Fig. 15. When the acceleration acts along the Y-axis, the maximum von-Mises stress of the structure is 494 MPa. According to the fourth strength theory, this value is much lower than the yield limit of carbon fiber reinforced polymer (CFRP) at 4000 MPa, verifying the load-bearing safety of the structure.

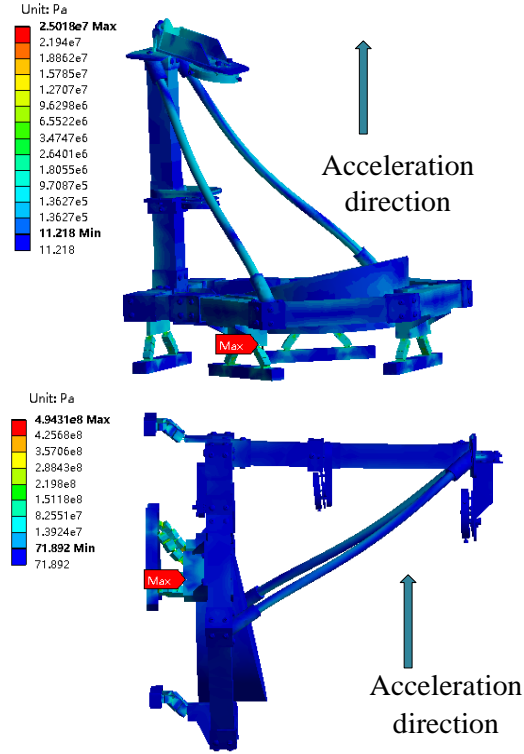

Fig. 15 Overall Machine Stress Cloud Diagram Under 10G Acceleration

**4.2 Temperature Stability Analysis**

During its on-orbit operation, the telescope will endure extreme and severe spatial temperature fluctuations. Such temperature variations may induce thermal stress in the structure, which further leads to distortion of the mirror surface and relative positional offset between mirrors. To ensure that the surface figure accuracy of the mirrors and the positional accuracy between mirrors still meet the design specifications under a temperature difference of 60 ℃, thermal stress simulation analysis is carried out on the entire telescope.

In the finite element analysis, a temperature load of 60 ℃ is applied to the overall structural model. The simulation results are shown in Fig. 16, and the specific deformation distribution of each mirror under this temperature difference is illustrated in Fig. 17. To quantitatively evaluate the influence of temperature on the optical system, the surface deformation data of each mirror are further extracted, and the variation of the distance between mirrors is calculated according to the formula, providing key data support for the subsequent judgment of whether the temperature adaptability meets the requirements.

$$\begin{cases} TZ = DZ(C) \\ TX = \dfrac{DX(A) \pm DX(B)}{2} \\ TY = \dfrac{DY(A) \pm DY(B)}{2} \\ TILT1 = \arctan(\dfrac{DZ1(A) - DZ1(B)}{D}) \\ TILT2 = \arctan(\dfrac{DZ2(A) - DZ2(B)}{D}) \\ TILT12 = TILT1 \pm TILT1 \end{cases} \quad (4)$$

where *TZ* is the axial displacement between the two mirrors, *TX* and *TY* are the displacements of the two mirrors along the X-axis and Y-axis directions respectively, *TILT1* and *TILT2* are the tilt angles of the two mirrors, and *TILT12* is the relative tilt angle between the two mirrors. In the formula, the operator is "−" when the displacement directions are the same, and "+" otherwise; it is "−" when the tilt directions are the same, and "+" otherwise.

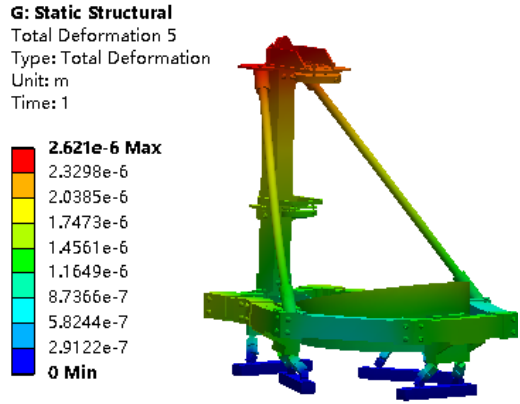

Fig. 16 Deformation Under a Temperature Change of 60℃

The deformation data of each surface are extracted, and the variations of inter-mirror parameters are calculated according to Equation 4, as listed in Table 3. It can be seen that the position variations meet the optical design specifications, verifying that the structure possesses good stability under such temperature changes.

Table 3 Relative Position Changes of Reflectors Under an Environmental Temperature Change of 60℃

| Mirrors | Axial Displacement/μm | Eccentricity /μm | Tilt /arcmin |
|---|---|---|---|
| PM-SM | 0.966 | X：0.8563<br>Y：0.2898 | X：≈0<br>Y：0.00015 |
| SM-TM | 1.5759 | X：0.2101<br>Y：0.1791 | X：≈0<br>Y：0.00003 |
| SM-FM | 0.6251 | X：0.1937<br>Y：0.3108 | X：≈0<br>Y：0.00003 |

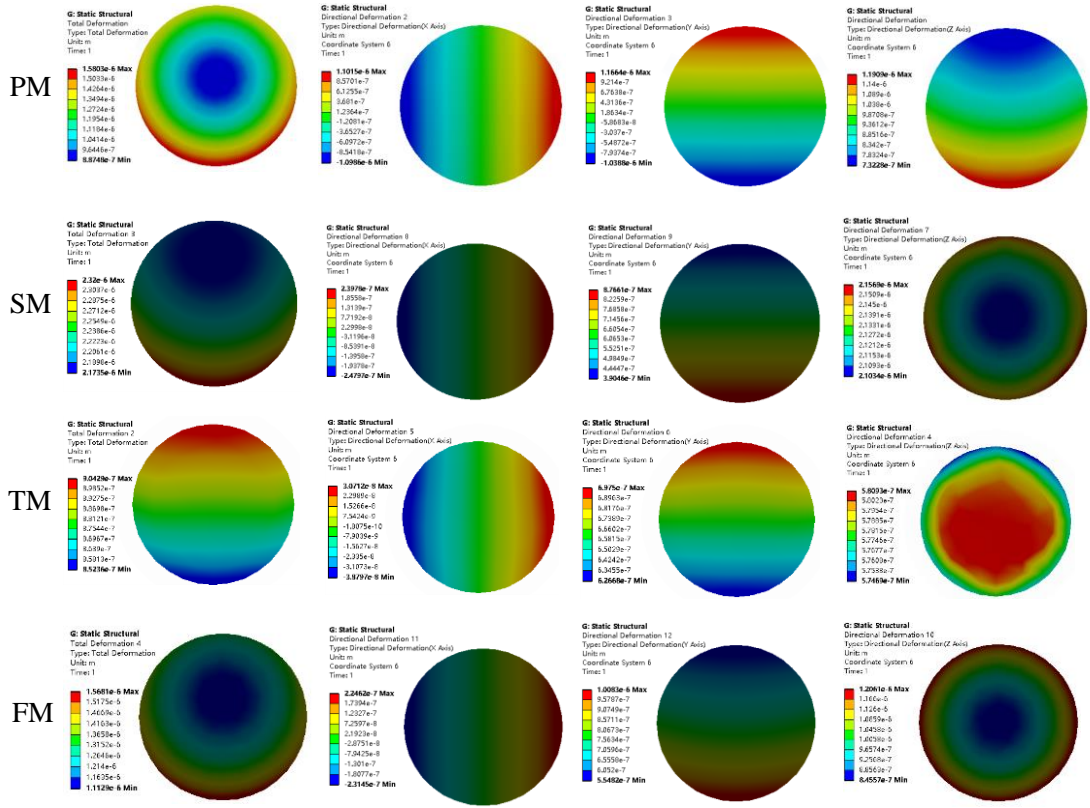

Fig. 17 Surface Deformation of Each Reflector Under a Temperature Change of 60℃

**4.3 Modal Analysis**

As a core indicator characterizing the dynamic stiffness of the system, modal analysis of the overall structure is critical for evaluating dynamic stability and avoiding resonance risks. In this paper, the minimum modal frequency of the structure is specified as 80 Hz. Considering the difference in gravitational acceleration between ground and space environments, both natural modal and prestressed modal analyses are performed on the entire telescope structure.

In the natural modal analysis, binding constraints are applied to the threaded connections of the mirrors, and fixed constraints are imposed on the threaded holes of the flexure hinges. In the prestressed modal analysis, gravitational acceleration is additionally applied based on the natural modal constraints to simulate the actual gravity environment.

Through the above analyses, the first six orders of natural mode shapes and prestressed mode shapes of the mirror are obtained, as shown in Fig. 18 and Fig. 19 respectively. The first six modal frequencies of the overall structure are listed in Table 4. It can be seen that the first-order natural frequency of the telescope structure is 221.209 Hz, which is higher than the design requirement of 80 Hz.

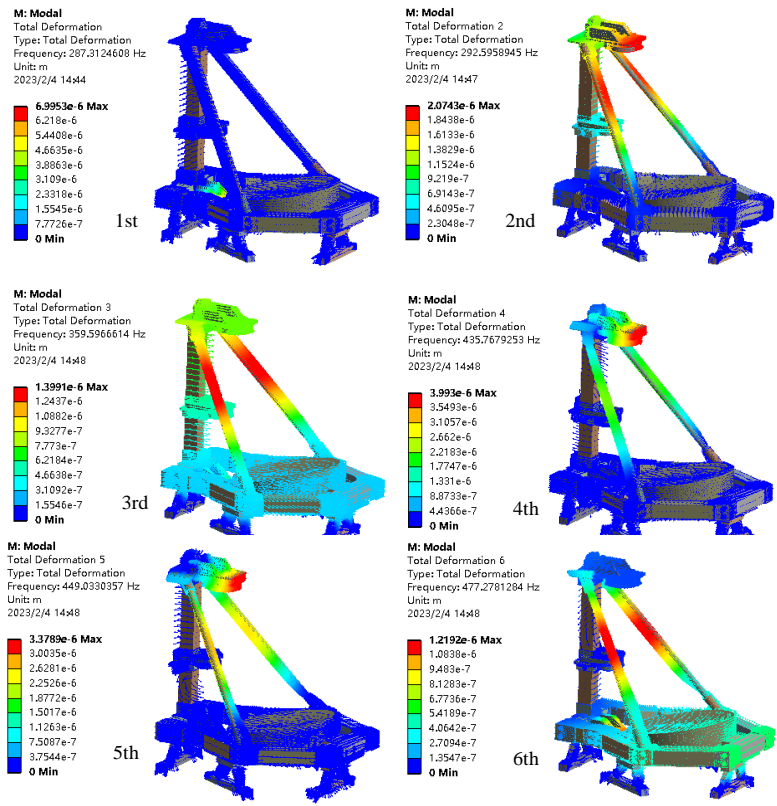

Fig. 17 Natural Modes of the Overall Structure

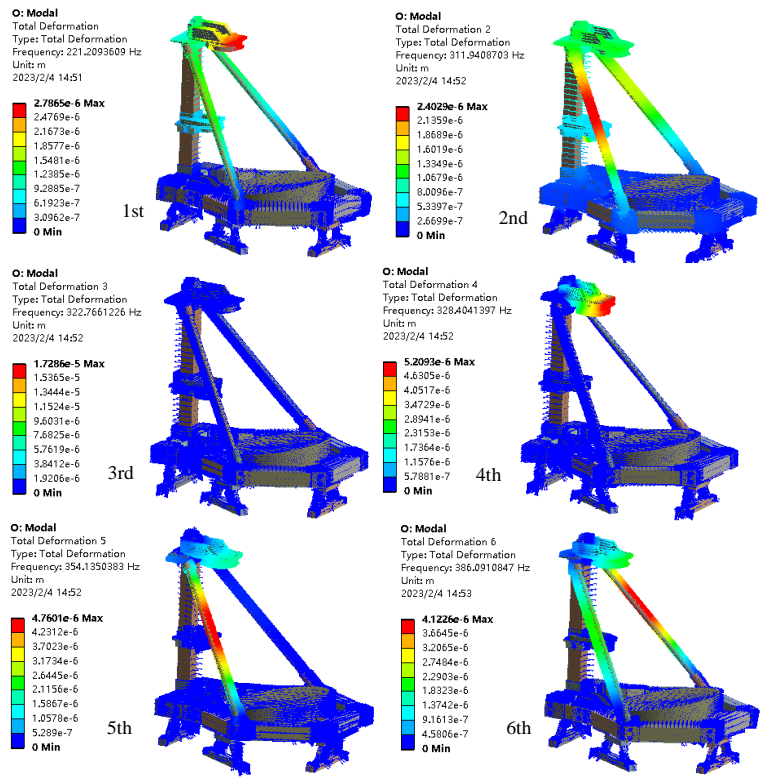

Fig. 18 Prestressed Modes of the Overall Structure

Table 4 First Six Orders of Modal Frequencies of the Overall Structure

| Order | 1st | 2nd | 3rd | 4th | 5th | 6th |
|---|---|---|---|---|---|---|
| Natural Mode Value/Hz | 287.312 | 292.596 | 359.597 | 435.768 | 449.03 | 477.278 |
| Prestressed Mode Value /Hz | 221.209 | 311.941 | 322.766 | 328.404 | 354.135 | 386.091 |

# 5  Conclusion

This paper conducts a systematic study on the structural design of a spaceborne laser transmitting telescope. In response to the space environment effects, the primary mirror adopts a flexible support combined with lightweight design to relieve environmental stress and reduce the influence of thermal deformation and vibration on the mirror surface. Precision fine-adjustment mechanisms are equipped for each mirror to ensure controllable spatial positioning accuracy. Meanwhile, multi-condition performance verification is carried out using finite element simulation: the overload resistance of the structure is evaluated under 10G acceleration simulating the launch process; the influence of thermal deformation on inter-mirror distance and surface figure accuracy is analyzed under extreme on-orbit temperature difference; the natural and prestressed modes with and without gravity load are compared to verify dynamic stability. Multi-dimensional simulation results show that the designed telescope structure can meet the preset optical design requirements under extreme mechanical loads, temperature variations, and dynamic working conditions, which fully validates the reliability and engineering applicability of the structural scheme.

**References：**